______________________________________________

# EFFECT OF SUBSTITUENTS IN POLYVINYLCARBAZOLE STRUCTURES ON THEIR OPTICAL PROPERTIES

Ya.Vertsimakha[1], P.Lutsyk[1], V.Syromyatnikov[2], I.Savchenko[2]

(1) Institute of Physics, NASU, prosp. Nauky 46, 03680, Kyiv, Ukraine
(2) Macromolecular Chemistry Department, Taras Shevchenko National University of Kyiv, 64 Volodymyrska str., 01601, Kyiv, Ukraine

E-mail: iras@univ.kiev.ua

Absorption, photoluminescence, and photoluminescence excitation spectra of solutions and thin films of N-vinylcarbazole polymers and copolymers with various substituents directly on the carbazole moiety and on the polymer chain were studied comprehensively. Polymers that were used previously to develop polymer composites with polymethine dyes having photosensitivity over a broad spectral range including the visible and near-IR regions were selected for the studies.

УДК 621.315.592:668.819.5

# ВЛИЯНИЕ ЗАМЕСТИТЕЛЕЙ В СТРУКТУРЕ ПОЛИВИНИЛКАРБАЗОЛОВ НА ИХ ОПТИЧЕСКИЕ СВОЙСТВА

*Я. Верцимаха[1], П.Луцик[1], В.Сыромятников[2], И.Савченко*[2]*

[1]Институт физики НАН Украины, Украина, 03680, Киев, проспект Науки, 46

[2] Киевский национальный университет имени Тараса Шевченко, Украина, 01601, Киев, ул. Владимирская, 64

E-mail: iras@univ.kiev.ua

*Впервые проведены комплексные исследования поглощения, фотолюминесценции (PL) и возбуждения фотолюминесценции (EPL) растворов и тонких пленок полимеров и сополимеров на основе N-винилкарбазола с различными заместителями непосредственно в карбазольном фрагменте и в полимерной цепи. Для исследования были выбраны полимеры, которые ранее использовались для разработки полимерных композитов с полиметиновыми красителями с высокой фоточувствительностью в широком спектральном диапазоне, включая видимую и часть ближней инфракрасной области.*

**Ключевые слова:** фотолюминесценция, возбуждение фотолюминесценции, спектры поглощения**,** N-винилкарбазол, сополимер, октилметакрилат, эксимер

**Введение.** Поли(N-винилкарбазол) был предметом интенсивного исследования в последние 50 лет с момента открытия его фотопроводимости Hoegl [1]. С того времени большое количество различных карбазолсодержащих полимеров было синтезировано и исследовано и при этом показано, что такие полимеры имеют хорошие фоторефрактивные, оптические свойства и свойства переноса заряда [2].

В последнее время существенно возрос интерес к их исследованию в связи с разработкой на основе их производных пластических композитов, фоточувствительных в широкой спектральной области [3-5], органических светодиодов [6], материалов для записи информации [7], органических солнечных элементов [8] и слоев в многослойных структурах [9].

Несмотря на это, электронные спектры поглощения слабо изучены. Особенно мало исследовано влияние заместителей в структуре полимеров на спектры возбуждения фотолюминесценции пленок производных этих соединений [10-11].

Поэтому целью данной работы было исследование влияния заместителей в структуре полимеров и сополимеров N-винилкарбазола на их оптические свойства для разработки методов дальнейшего повышения фоточувствительности этих полимерных композитов с красителями.

## 2. Эксперимент

Растворимые в легко летучих растворителях полимеры и сополимеры N-винилкарбазола с октилметакрилатом (VC-OMA), N-винил-3-йодкарбазола с октилметакрилатом (I-VC-OMA), N-винил-7H-бензокарбазол (V7BC) и его сополимер с ОМА (V7BC-OMA) (схема 1) были синтезированы и очищены перекристаллизацией из соответствующих органических растворителей на кафедре химии высокомолекулярных соединений Киевского национального университета имени Тараса Шевченко имеют следующую структуру:

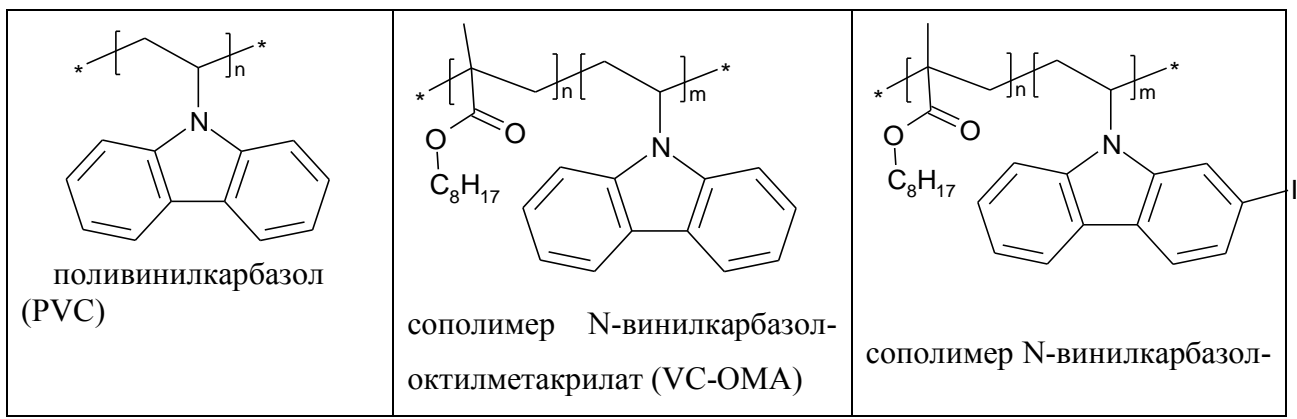

поливинилкарбазол (PVC)

сополимер N-винилкарбазол-октилметакрилат (VC-OMA)

сополимер N-винилкарбазол-

| | | 3-йодоктилметакрилат (I-VC-OMA) |
|---|---|---|
| 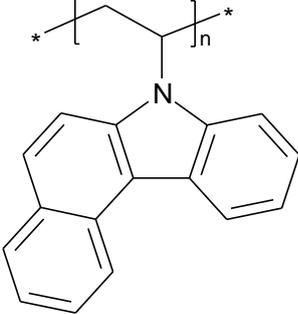 поли(N-винил-7H-бензо(c)карбазол) (V7BC) | 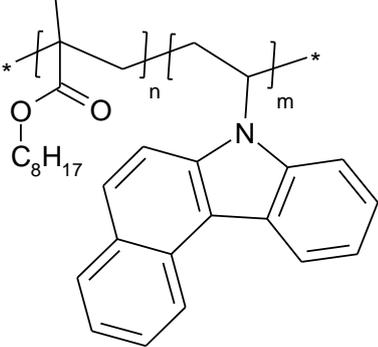 сополимер N-винил-7H-бензо(c)карбазол-OMA (V7BC-OMA) | |

Схема 1. Структуры исследуемых полимеров.

Полимеризацию N-винилкарбазола, V7BC и сополимеризацию VC:OMA, V7BC-OMA 3-I-VC-OMA проводили в присутствии азобисизобутиронитрила (АІБН) как инициатора в растворе пропанола-2 или толуола при 80$^\circ$C в атмосфере аргона на протяжении 1-2 час. Полимеры высаживали в метанол. Очистку полимеров проводили перекристаллизацией из бензола (или диоксана) в метанол.

Тонкие пленки полимеров толщиной 30-100 нм были получены поливом раствора полимера в хлороформе на кварцевые подкладки.

Спектры поглощения исследуемых растворов и пленок измерялись с помощью спектрофотометров Shimadzu UV2450 и UV-VIS, для измерения спектров фотолюминесценции и ее возбуждения применялся спектрофотометр Jobin Yvon NanoLog.

**3. Результаты и их обсуждение**

По данным [11] в спектре поглощения раствора PVC в гексане в области 3-4 эВ проявляются π-π* переходы $^1A{\to}^1L_b$ и $^1A{\to}^1L_a$. Наши измерения спектров поглощения растворов в хлороформе и пленок исследуемых полимеров, показали, что изменение агрегатного состояния (переход от раствора к пленке) приводит к небольшому красному сдвигу (8-10 мэВ) максимумов полос поглощения без изменения соотношения интенсивности их полос поглощения. Это свидетельствует о слабом взаимодействии молекул полимеров в пленках всех исследуемых полимеров и сополимеров и поэтому для простоты ниже на рисунках будут приводиться спектры поглощения растворов (точность измерения энергетического положения максимумов полос 5 мэВ).

Комплексные измерения поглощения и фотолюминесценции пленок PVC показывают, что максимумы поглощения $^1A \rightarrow ^1L_b$ перехода и возбуждения люминесценции совпадают (рис.1), и сигнал EPL незначительный в области hv > 3.8 эВ (в области второго электронного перехода), то есть излучение люминесценции происходит из нижнего возбужденного электронного уровня [10]. Большая ширина полосы фотолюминесценции (больше 1 эВ) с максимумом при 3.10 эВ, ее асимметрия  и наличие перегиба около 3.26 эВ (рис.1) свидетельствуют о том, что она состоит из нескольких полос . Эти полосы фотолюминесценции более четко видны в спектрах фотолюминесценции пленок PVC-OMA и V7BC-OMA (рис.2 и рис.3). Последнее связано с большей пластичностью PVC-OMA и V7BC-OMA, что приводит к уменьшению количества дефектов в их пленках.

Рис.1

Большая величина стоксового сдвига (0.5 эВ) полосы с максимумом при 3.10 эВ (таблица 1) и отсутствие поглощения в пленках в этой спектральной области свидетельствуют, что эта полоса имеет эксимерную природу и большую вероятность образования эксимеров при формировании пленки PVC. Мономерная полоса флуоресценции вероятно находится в области 3.1-3.3 эВ [10], а ее интенсивность в исследуемых нами пленках PVC значительно ослаблена.

Таблица 1

При сравнении спектров поглощения пленок PVC (рис.1)  и VC-OMA (рис.2) видно, что введение гибких цепочек ОМА путем сополимеризации с N-винилкарбазолом приводит к небольшому смещению максимумов полос VC-OMA в УФ область, до 10 мэВ (таблица 1, рис.2). что может быть обусловлено незначительным уменьшением межмолекулярного взаимодействия. Максимум EPL практически совпадает с максимумом полосы поглощения $^1A \rightarrow ^1L_b$  перехода, т.е. излучение фотолюминесценции происходит со дна возбужденной зоны $^1A \rightarrow ^1L_b$ перехода.

**Рис.2**

Влияние ОМА более четко  проявляется в спектрах фотолюминесценции (рис.2). Наблюдается увеличение структурности спектра фотолюминесценции (уже проявляются две полосы с максимумом при 3.17 и 3.34 эВ и перегиб около 3.55 эВ) (рис.2, таблица 1). Наиболее вероятно, что увеличение структурности полосы PL обусловлено уменьшением ширины полос эксимерного свечения вследствие улучшения структуры полимерных агрегатов и уменьшения рассеивания света на структурных дефектах в пленках. Это четко наблюдается даже при визуальном сравнении морфологии пленок PVC и VC-OMA. Пленки VC-OMA более однородны и меньше рассеивают падающий свет. Это можно объяснить увеличением пластичности пленок и экспериментально подтверждается увеличением растворимости VC-OMA в органических растворителях.

Однако введение йода в молекулу N-винилкарбазола приводит к значительному (80 мэВ) смещению полос поглощения в длинноволновую область и значительному (200 мэВ) смещению максимума возбуждения фотолюминесценции в коротковолновую область (таблица 1). Это можно объяснить увеличением эффективности излучения молекулами, которые поляризованы вдоль их a-оси. Одновременно возрастает количество структурных дефектов и рассеивания света на них, что приводит к размыванию спектра фотолюминесценции (проявляются два максимума  рис.2). Вследствие этого, форма полосы PL I-VC-OMA похожа на форму полосы PL PVC (рис.1), но энергия максимумов полос фотолюминесценции смещена на 70 мэВ в сторону больших энергий (таблица 1).

Формирование молекулы V7BC вследствие введения бензольного ядра в молекулу VC приводит к существенному возрастанию величины смещения энергии $\Delta E$ самого нижнего электронного $^1A{\rightarrow}^1L_b$ перехода в длинноволновую область (до 250 мэВ) в сравнении со смещением при введением йода в молекулу VC (80мэВ)  и возрастанию интенсивности поглощения (вероятности) $^1A{\rightarrow}^1L_a$ перехода при 3.38 эВ, которая становится практически в 2 раза больше интенсивности $^1A{\rightarrow}^1L_b$  (рис. 3).

**Рис.3**

При этом максимум полосы возбуждения фотолюминесцении пленок V7BC практически совпадает с максимумом $^1A \rightarrow ^1L_a$ перехода, то есть излучение люминесценции происходит уже практически со дна возбужденной электронной зоны $^1A \rightarrow ^1L_a$ перехода.

Вероятно, при формировании пленки V7BC возрастает число структурных дефектов, вследсвие чего в спектре фотолюминесцении явно проявляются только два перегиба на кривой PL области эксимерной люминесценции - при 2.56 и 3.0 (рис.3). Вид спектров фотолюминесцении I-VC-ОМА и V7BC похож. Однако энергия эксимеров в V7BC на 0.3 эВ меньше, чем эксимеров в I-VC-ОМА, вследствие уменьшения энергии взаимодействия между молекулами эксимеров в V7BC.

Несмотря на высокую вероятность формирования структурных дефектов из молекул V7BC, при его сополимеризации с ОМА, который содержит гибкие спенсеры, уменьшаются стерические затруднения для формирования в пленках V7BC-ОМА эксимеров, вследствие чего в их спектрах фотолюминесцении проявляются уже 2 максимума и два явных перегиба (рис.3). Вероятно, что симметрия этих эксимеров в пленках существенно отличается, так как их излучение происходит как со дна возбужденной зоны $^1A \rightarrow ^1L_b$, так и $^1A \rightarrow ^1L_a$ переходов. Это подтверждается сложным спектром возбуждения фотолюминесцении пленок V7BC - ОМА (рис.3) и увеличением  величины стоксова сдвига (таблица 2), которая зависит от поляризации электронного перехода.

 **Таблица 2**

Последнее показывает улучшение молекулярной упорядоченности в пленках с длинными спейсерами ОМА и заместителями - йодом в I-VC-ОМА и бензольным ядром в (V7BC-ОМА) в сравнении с аналогами (PVC и V7BC) без ОМА, что улучшает структуру полимерных агрегатов в пленках.

**Заключение.** Установлено, щo введение в молекулу PVC путем сополимеризации гибких спейсеров типа ОМА существенно увеличивает растворимость и упорядоченность полученных пленок PVC в сравнении с пленками без ОМА. Это приводит к небольшому сужению полос поглощения и возбуждения фотолюминесцении.  Наличие гибких спейсеров ОМА в полимерной цепи PVC и V7BC влияет на энергию максимумов полос эксимерной флуоресценции. Вследствие чего, например, в пленках VC-ОМА наблюдаются максимумы полосы эксимерной PL при 3.34 и 3.17 эВ, которые в пленках PVC без ОМА перекрывались и наблюдалась только широкая полоса фотолюминесценции с максимумом при 3.10 эВ.

Изменения во всех спектрах вследствие введения заместителей в молекулу PVC сильно зависят от размера (объема) заместителей. Так, введение йода приводит, главным образом, к значительному красному сдвигу (77 мэВ) $^1A{\rightarrow}^1L_b$ электронного перехода поляризованных вдоль короткой оси молекул полимера. А введение бензольного ядра в структуру винилкарбазола приводит к существенному увеличению интенсивности $^1A{\rightarrow}^1L_b$ переходов, поляризованных вдоль короткой оси молекулы винилкарбазола, и к большему красному сдвигу максимума (250 мэВ) самого нижнего по энергии $^1A{\rightarrow}^1L_b$ переходов. Кроме этого, эта замена приводит к голубому сдвигу энергетического положения максимума спектров EPL в область поглощения $^1A{\rightarrow}^1L_b$ переходов (на 530 мэВ).

Полученные данные могут быть использованы для разработки новых методов дальнейшего повышения фоточувствительности полимерных композитных материалов в широком спектральном диапазоне.

## Список литературы

**Подписи к рисункам.**

Рис.1. Спектральные зависимости поглощения раствора PVC (1-D, масштаб справа), возбуждения люминесценции (2- EPL) и люминесценции(3-PL) пленок PVC.

Рис.2. Спектральные зависимости поглощения раствора (1,4 - масштаб справа), возбуждения люминесценции (2, 5 - EPL) и люминесценции (3, 6 - PL) пленок VC-OMA (1-3) и I-VC-OMA (4- 6).

Рис.3. Спектральные зависимости поглощения растворов (1,4 - масштаб справа), возбуждения фотолюминесценции (2,5 - EPL) и фотолюминесценции (3, 6 - PL) пленок V7BC (1-3) и V7BC-OMA (4-6).

**Таблица 1. Максимумы полос в спектрах поглощения D, фотолюминесценции PL и ее возбуждения EPL пленок карбазолсодержащих полимеров и их сдвиг ΔЕ относительно соответствующих полос PVC (в скобках)**

| Полимеры→ | PVC | VC-OMA | I-VC-OMA | V7BC | V7BC-OMA |
|---|---|---|---|---|---|
| Спектры ↓ | Емах, эВ | Емах(ΔЕ), эВ | Емах(ΔЕ), эВ | Емах(ΔЕ), эВ | Емах(ΔЕ), эВ |
| В спектрах поглощения D | 3.60<br>3.76<br>3.89 | 3.61(+0.01)<br>3.77 (+0.01)<br>- | 3.52(-0.08)<br>3.67 (-0.10)<br>3.81(-0.08) | 3.38(-0.22)<br>3.55 (-021)<br>- | 3.35(-0.25)<br>3.53(-0.23)<br>3.77(-0.12) |
| В спектрах возбуждения фотолюми-несценции | 3.58 | 3.56(-0.02) | 3.50*‡<br>3.78 (+0.2) | -<br>-<br>3.78 (+0.20) | 3.35(-0.23)<br>3.52*<br>3.69 * |
| В спектрах фотолюминес ценции | 3.26 ‡<br>3.10 | 3.55* ‡<br>3.34 (+0.08)<br>3.17 (+0.07) | 3.32(+0.06) ‡<br>3.17(+0.07) | 3.0(-0.26)<br>2.56(-0.54) ‡ | 3.30*<br>3.14(-0.12)<br>2.97(-0.13) |
| * не наблюдается соответствующих полос в спектрах PVC<br>‡ энергия полос в области перегибов спектра | | | | | |

3.37* в EPL VC-OM  - а чому це її немає в EPL PVC?

**Таблица 2. Енергетичне положення Eмах полос поглощения и фотолюминесценции, их интенсивность относительно наиболее интенсивной полосы I/I₁ и величина стоксового сдвига для пленок полимеров и сополимеров VC, нанесенных на кварцевую подложку из их растворов в хлороформе.**

| Пара метры | PVC | | VC-OMA | | I-VC-OMA | | V7BC | | V7BC-OMA | |
|---|---|---|---|---|---|---|---|---|---|---|
| | Eмах (эВ) | $I/I_1$ | Eмах (эВ) | $I/I_1$ | Eмах (эВ) | $I/I_1$ | Eмах (эВ) | $I/I_1$ | Eмах (эВ) | $I/I_1$ |
| Поглощение | 3.60 | 1.00 | 3.61 | 1.0 | 3.52 | 1.00 | 3.38 | 0.35 | 3.35 | 0.39 |
| Фото- люминесце нция | | | **3.55** | 0.25 | | | | | 3.30 | 0.56 |
| | 3.26 | 0.70 | **3.34** | 1.00 | 3.32 | **0.88** | 3.0 | 1.0 | 3.14 | 1.00 |
| | 3.10 | 1.00 | **3.17** | 0.99 | 3.17 | **1.00** | 2.56 | 0.20 | 2.97 | 0.73 |
| Стоксов сдвиг, эВ | | | 0.06 | | | | | | 0.05 | |
| | 0.34 | | 0.27 | | 0.21 | | 0.38 | | 0.21 | |
| | 0.50 | | 0.44 | | 0.35 | | 0.82 | | 0.38 | |

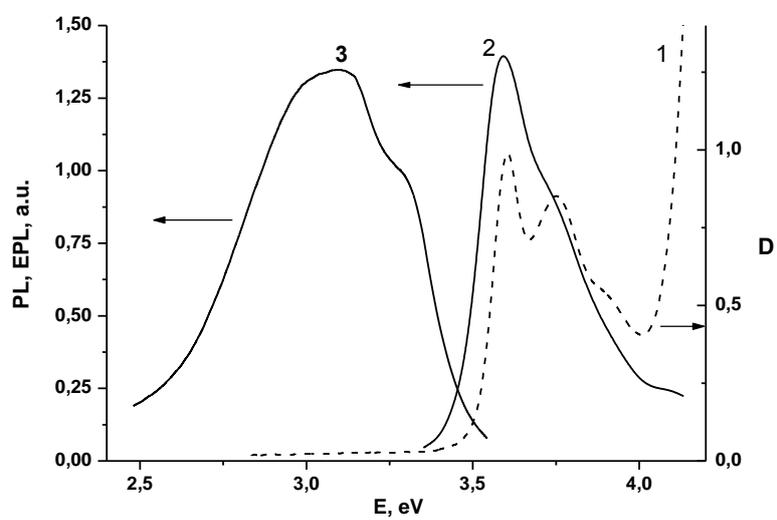

Рис.1

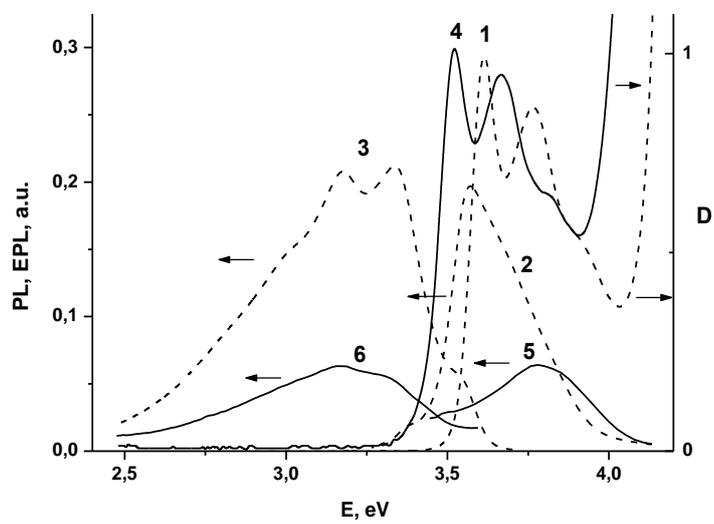

Рис.2.

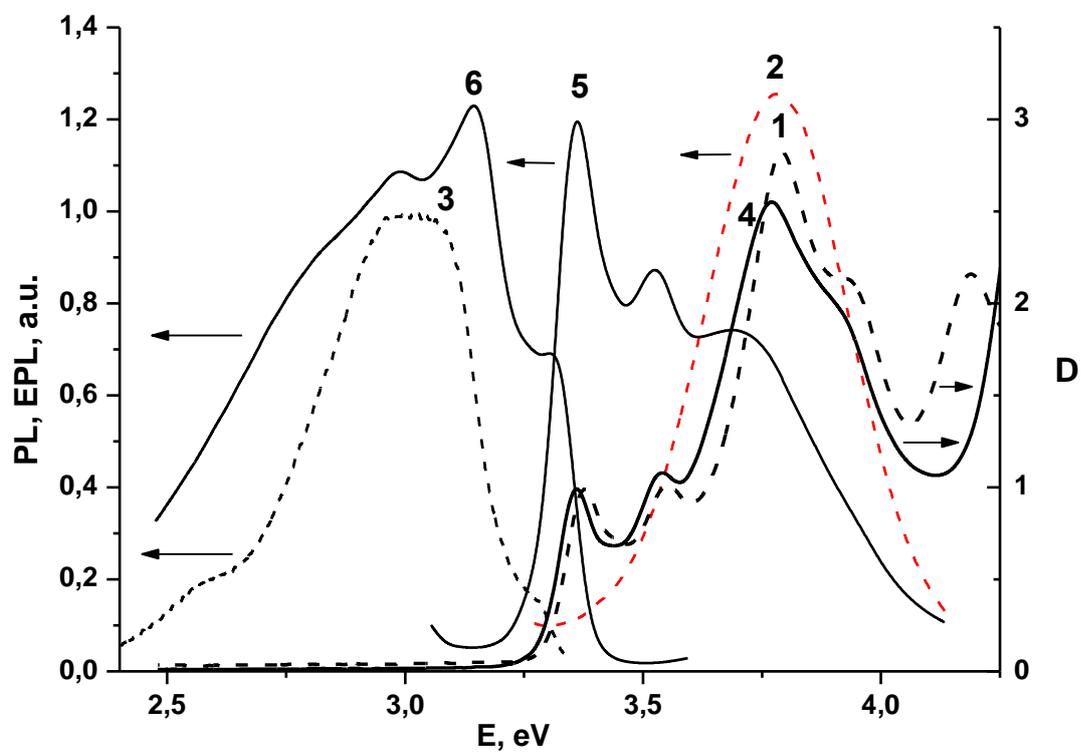

Рис.3.